\input{genfu3nj.def}
\documentstyle[11pt,epsf]{article}\headsep0cm
\begin{document} 
\title{Generating Functions for Multi-$j$-Symbols} 
\author{Oliver Schnetz 
\thanks{DIAS, 10 Burlington Rd., Dubin 4, Ireland\newline 
\hspace*{4ex}e-mail: schnetz@online.de}} 
\maketitle 
\begin{abstract} 
A formula is derived that provides generating functions for any multi-$j$-symbol, 
such as the 3-$j$-symbol, the 6-$j$-symbol, the 9-$j$-symbol, etc. 
The result is completely determined by geometrical objects (loops 
and curves) in the graph of the multi-$j$-symbol. A geometric-combinatorical 
interpretation for multi-$j$-symbols is given. 
\end{abstract} 
\tableofcontents 
\section{Introduction} 
The 3-$j$-symbol describes how the tensor product of two $ SU(2)
$ representations $ R^{(j_{1})}$, $ R^{(j_{2})}$ with dimensions 
$ 2j_{1}+1$, $ 2j_{2}+1$ can be reduced to a direct sum of (the 
complex conjugate of) representations $ R^{(j_{3})}$* \cite{A}: 
\begin{equation}
\label{15}R^{\left( j_{1}\right) }_{m_{1}m_{1}'}\left( x\right) R^
{\left( j_{2}\right) }_{m_{2}m_{2}'}\left( x\right) =\sum _{j_{
3}m_{3}m_{3}'}\left( 2j_{3}+1\right) \left( 
\begin{array}{ccc}j_{1}&j_{2}&j_{3}{}\\ 
m_{1}&m_{2}&m_{3}
\end{array}\right) R^{\left( j_{3}\right) *}_{m_{3}m_{3}'}
\left( x\right) \left( 
\begin{array}{ccc}j_{1}&j_{2}&j_{3}{}\\ 
m_{1}'&m_{2}'&m_{3}'
\end{array}\right) \hspace{.6ex},
\end{equation} 
with $ x\in SU(2)$. The $j$$_{1}$, $j$$_{2}$, $j$$_{3}$ are half 
integers and $ m_{1}=-j_{1}$, $ -j_{1}+1$, {\dots}, $j$$_{1}$, etc. 
Obviously Eq.\ (\ref{15}) depends on the explicit choice of the 
representations $ R^{(j)}$. As usual we use Euler parameters to 
parameterize $ SU(2)$ and choose representations that are diagonal 
in the first and the third rotation: 
\begin{equation}
\label{18}R^{j}_{mm'}\left( \alpha ,\beta ,\gamma \right) =\exp
\left( im\alpha +im'\gamma \right) d^{j}_{mm'}\left( \beta 
\right) \hspace{.6ex},
\end{equation} 
where the $ d^{j}_{mm'}$ are related to the Jacobi polynomials by 
\begin{equation}
d^{j}_{mm'}\left( \beta \right) =\sqrt\frac{\left( j+m\right) !
\left( j-m\right) !}{\left( j+m'\right) !\left( j-m'\right) !} 
\left( \cos \frac{\beta }{2} \right) ^{m+m'}\left( \sin \frac{
\beta }{2} \right) ^{m-m'}P^{\left( m-m',m+m'\right) }_{j-m}
\left( \cos\left( \beta \right) \right) \hspace{.6ex}.
\end{equation} 
For the 3-$j$-symbol a generating function is well known \cite{B}: 
\begin{eqnarray}
\label{7}&&\sum _{\alpha \beta \gamma }\left( 
\begin{array}{ccc}a&b&c\\ 
\alpha &\beta &\gamma 
\end{array}\right) \sqrt\frac{\left( a+b+c+1\right) !\left( a+b-c
\right) !\left( b+c-a\right) !\left( c+a-b\right) !}{\left( a+
\alpha \right) !\left( a-\alpha \right) !\left( b+\beta 
\right) !\left( b-\beta \right) !\left( c+\gamma \right) !
\left( c-\gamma \right) !} \cdot \\ 
&&\cdot A^{a+\alpha }\bar{A}^{a-\alpha }B^{b+\beta }\bar{B}^{b-
\beta }C^{c+\gamma }\bar{C}^{c-\gamma }=\left( A\bar{B}-B\bar
{A}\right) ^{a+b-c}\left( B\bar{C}-C\bar{B}\right) ^{b+c-a}
\left( C\bar{A}-A\bar{C}\right) ^{c+a-b}\hspace{.6ex},
\nonumber 
\end{eqnarray} 
where $ A$, $ \bar{A}$, $ B$, $ \bar{B}$, $ C$, $ \bar{C}$ are all 
independent parameters.

If we divide both sides of this equation by $ (a+b-c)!(b+c-a)!(c
+a-b)!$, sum over $a$, $b$, and $c$, and introduce the shorthand 
\begin{equation}
\label{9}\Delta \left( a,b,c\right) =\left( \frac{\left( a+b-c
\right) !\left( b+c-a\right) !\left( c+a-b\right) !}{\left( a+b
+c+1\right) !} \right) ^{\frac{1}{2}}
\end{equation} 
we find 
\begin{eqnarray}
&&\sum _{\textstyle  {abc\atop \alpha \beta \gamma }}\left( 
\begin{array}{ccc}a&b&c\\ 
\alpha &\beta &\gamma 
\end{array}\right) \frac{A^{a+\alpha }\bar{A}^{a-\alpha }B^{b+
\beta }\bar{B}^{b-\beta }C^{c+\gamma }\bar{C}^{c-\gamma }}{
\Delta \left( a,b,c\right) \sqrt{\left( a+\alpha \right) !
\left( a-\alpha \right) !\left( b+\beta \right) !\left( b-
\beta \right) !\left( c+\gamma \right) !\left( c-\gamma 
\right) !}}\nonumber \\ 
\label{1}&&=\sum _{a+b+c=0}^{\infty }\frac{\left( A\bar{B}-B\bar
{A}+B\bar{C}-C\bar{B}+C\bar{A}-A\bar{C}\right) ^{a+b+c}}{
\left( a+b+c\right) !} =\exp\left( \left| 
\begin{array}{ccc}1&A&\bar{A}\\ 
1&B&\bar{B}\\ 
1&C&\bar{C}
\end{array}\right| \right) \hspace{.6ex}.
\end{eqnarray} 
Note that we have six expansion parameters $ A$, $ \bar{A}$, $ B
$, $ \bar{B}$, $ C$, $ \bar{C}$ for six variables $a$, $\alpha 
$, $b$, $ \beta $, $c$, $\gamma $. We do not lose access to the 
3-$j$-symbol by the extra sums over $a$, $b$, $c$; the 3-$j$-symbol 
can be extracted by expanding the right hand side of Eq.\ (\ref{1}). 
This equation reveals the full symmetry structure of the 3-$j$-symbol 
including the Regge symmetries \cite{B} $ a\pm \alpha 
\rightarrow b+c-a$, $ b\pm \beta \rightarrow c+a-b$, $ c\pm 
\gamma \rightarrow a+b-c$, corresponding to $ A\rightarrow A
\sqrt{\bar{B}\bar{C}/\bar{A}}$, $ \bar{A}\rightarrow \sqrt{\bar
{B}\bar{C}/\bar{A}}$ plus cyclic permutations or $ A
\rightarrow \sqrt{\bar{B}\bar{C}/\bar{A}}$, $ \bar{A}
\rightarrow \bar{A}\sqrt{\bar{B}\bar{C}/\bar{A}}$ (+ cycl.), respectively.

Of course, Eq.\ (\ref{1}) is not a generating function of the 3-$j$-symbol 
itself, but a generating function of the 3-$j$-symbol with a suitable 
normalization. This normalization eliminates the square roots in 
the 3-$j$-symbol and, obviously, without the normalization calculating 
a generating function would not be realistic.

\vspace{2ex}
\noindent{}If we substitute $ \bar{A}\rightarrow t_{1}\bar{A}$, $ 
\bar{B}\rightarrow t_{2}\bar{B}$, $ \bar{C}\rightarrow t_{3}
\bar{C}$ in Eq.\ (\ref{1}), multiply both sides by $ \exp(-t_{1
}-t_{2}-t_{3})$, and integrate over $ t_{1}$, $ t_{2}$, $ t_{3}
$ from 0 to $\infty $ we obtain another, equivalent generating function 
for the 3-$j$-symbol: 
\begin{eqnarray}
&&\sum _{\textstyle  {abc\atop \alpha \beta \gamma }}\left( 
\begin{array}{ccc}a&b&c\\ 
\alpha &\beta &\gamma 
\end{array}\right) \sqrt\frac{\left( a-\alpha \right) !\left( b-
\beta \right) !\left( c-\gamma \right) !}{\left( a+\alpha 
\right) !\left( b+\beta \right) !\left( c+\gamma \right) !} 
\frac{A^{a+\alpha }\bar{A}^{a-\alpha }B^{b+\beta }\bar{B}^{b-
\beta }C^{c+\gamma }\bar{C}^{c-\gamma }}{\Delta \left( a,b,c
\right) }\nonumber \\ 
\label{14}&&=\frac{1}{1+\left( B-C\right) \bar{A}} \hspace*{1ex}
\frac{1}{1+\left( C-A\right) \bar{B}} \hspace*{1ex}\frac{1}{1+
\left( A-B\right) \bar{C}} \hspace{.6ex}.
\end{eqnarray}

\vspace{1ex}
\noindent{}In the same spirit we also find generating functions for 
the 6-$j$-symbol and the 9-$j$-symbol: 
\begin{eqnarray}
\hspace{-.7cm}&&\sum _{\textstyle  {abc\atop def}}\left\{ 
\begin{array}{ccc}a&b&c\\ 
d&e&f
\end{array}\right\} \frac{A^{2a}B^{2b}C^{2c}D^{2d}E^{2e}F^{2f}}{
\Delta \left( a,b,c\right) \Delta \left( a,e,f\right) \Delta 
\left( c,d,e\right) \Delta \left( b,d,f\right) }\nonumber \\ 
\label{2}\hspace{-.7cm}&&=\left( 1+ABF+ACE+BCD+DEF+ABDE+ACDF+BCEF
\right) ^{-2}\hspace{.6ex},\\ 
\hspace{-.7cm}&&\sum _{\textstyle  {\textstyle  {abc\atop def}
\atop ghk}}\left\{ 
\begin{array}{ccc}a&b&c\\ 
d&e&f\\ 
g&h&k
\end{array}\right\} \frac{A^{2a}B^{2b}C^{2c}D^{2d}E^{2e}F^{2f}G^
{2g}H^{2h}K^{2k}}{\Delta \left( a,b,c\right) \Delta \left( d,e,f
\right) \Delta \left( g,h,k\right) \Delta \left( a,d,g\right) 
\Delta \left( b,e,h\right) \Delta \left( c,f,k\right) }
\nonumber \\ 
\hspace{-.7cm}\label{3}&&=\left( 1\!-\!ABDE\!-\!ABGH\!-\!ACDF\!-
\!ACGK\!-\!BCEF\!-\!BCHK\!-\!DEGH\!-\!DFGK\!-\!EFHK+\right. 
\nonumber \\ 
\hspace{-.7cm}&&\left. +ABEFGK\!+\!ACDEHK\!+\!BCDFGH\!-\!ABDFHK
\!-\!ACEFGH\!-\!BCDEGK\right) ^{-2}\hspace{.6ex}.
\end{eqnarray}

\vspace{1ex}
\noindent{}In the next section we will find a simple geometric interpretation 
of these formulae. In Sec.\ \ref{s3} we will prove a theorem that 
encorporates Eqs. (\ref{14}), (\ref{2}), (\ref{3}). We will find 
that by simple geometrical means it is possible to find a generating 
function for any multi-$j$-symbol.

\section{Graphical Notation} 
\subsection{Graphical Notation for Multi-$j$-Symbols} 
A 3-$j$-symbol is denoted by a three-valent vertex where the external 
lines carry the indices ($a$, $\alpha $), ($b$, $ \beta $), ($c$, 
$\gamma $) of the 3-$j$-symbol (or ($A$, $ \bar{A}$), ($B$, $ 
\bar{B}$), ($C$, $ \bar{C}$) for the generating function):

\vspace{1ex}
\noindent{}\centerline{\epsfbox{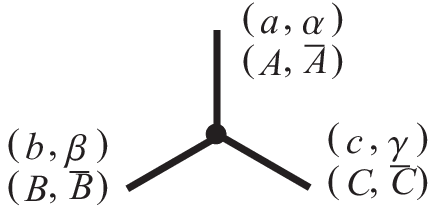}}

\vspace{1ex}
\noindent{}Note that the 3-$j$-symbol may pick up a minus sign under 
odd permutations of the columns. To keep track of those signs we 
implement the rule that the legs are labeled in a counter-clockwise 
orientation. Permuting to legs of the 3-$j$-symbol amounts to a 
factor $ (-1)^{a+b+c}$, which means on the level of generating functions 
$ A,B,C\rightarrow -A,-B,-C$, or, since $ \alpha +\beta +
\gamma =0$, equivalently $ \bar{A},\bar{B},\bar{C}\rightarrow - 
\bar{A},- \bar{B},- \bar{C}$.

\vspace{1ex}
\noindent{}Two external lines ($ a_{1}$, $ \alpha _{1}$) and ($ a
_{2}$, $ \alpha _{2}$) may be glued together with the group invariant 
'metric' 
\begin{equation}
\label{10}{a_{1}\choose \alpha _{1},\alpha _{2}}\delta _{a_{12},a
_{1}}\delta _{a_{12},a_{2}}=\left( -1\right) ^{a_{1}+\alpha _{1
}}\delta _{\alpha _{1},-\alpha _{2}}\delta _{a_{12},a_{1}}
\delta _{a_{12},a_{2}}
\end{equation}

\vspace{1ex}
\noindent{}\centerline{\epsfbox{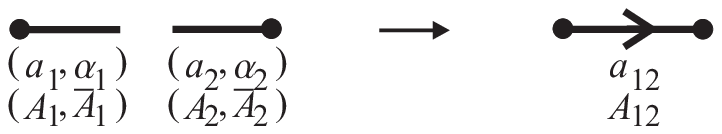}}

\vspace{1ex}
\noindent{}and sums over $ a_{1}$, $ a_{2}$, $ \alpha _{1}$, and 
$ \alpha _{2}$. The glued line has the angular momentum $ a_{12
}$ and no magnetic quantum number. The metric is not symmetric under 
exchange of ($ a_{1}$, $ \alpha _{1}$) and ($ a_{2}$, $ \alpha _{
2}$). It thus has an orientation which is denoted by the arrow. 
Changing the orientation of the arrow amounts in a factor $ (-1
)^{2a_{12}}$ or, on the level of the generating function ($ 
\sum _{a_{12}}f(a_{12})A_{12}^{2a_{12}}$), to $ A_{12}
\rightarrow -A_{12}$. 

\vspace{1ex}
\noindent{}With these rules every multi-$j$-symbol translates into 
a three-valent graph which we will call $\Gamma $. Every line in 
$\Gamma $ has an angular momentum quantum number (a $j$ variable, 
here labeled $a$, $b$, $c$, {\dots}, or $a$$_{1}$, $a$$_{2}$, $a$
$_{3}$, {\dots}). In addition, every external line has a magnetic 
quantum number (an $m$ variable, here labeled $\alpha $, $ 
\beta $, $\gamma $, {\dots}, or $\alpha _{1}$, $\alpha _{2}$, $
\alpha _{3}$, {\dots}). If there are no external lines the multi-$j$-symbol 
is closed, it has no magnetic quantum numbers. Closed multi-$j$-symbols 
have a group theoretical meaning independent of the specific representations 
chosen (Eq.\ (\ref{18})). However, we do not restrict ourselves 
to this case.

The standard 6-$j$-, and 9-$j$-symbols have the graphs:

\vspace{1ex}
\noindent{}\centerline{\epsfbox{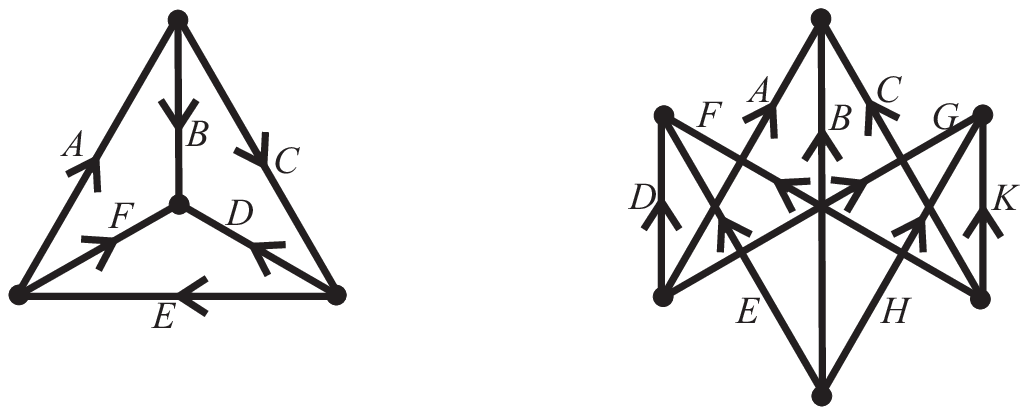}}

\vspace{1ex}
\noindent{}These are the most interesting however by no means the 
only symbols with six or nine angular momenta.

\subsection{Graphical Notation for the Generating Functions} 
Now we translate the quantum numbers $a$, $\alpha $ into expansion 
coefficients $A$, $ \bar{A}$ according to the following rules:

\vspace{1ex}
\noindent{}For internal lines we construct the generating function 
by multiplying with $ A^{2a}$ and summing over $ 2a=0,1,2,
{\ldots}$ . This means we change variables from small letters to 
capitals.

\vspace{1ex}
\noindent{}For external lines we multiply with $ A^{a+\alpha }$, 
$ \bar{A}^{a-\alpha }$ and sum over $ a+\alpha =0,1,2,{\ldots}$ 
and $ a-\alpha =0,1,2,{\ldots}$ (which is equivalent to $ a=0,
\frac{1}{2},1,{\ldots}$, $ \alpha =-a,-a+1,{\ldots},a$). We thus 
transform from ($a$, $\alpha $) to ($A$, $ \bar{A}$).

\vspace{1ex}
\noindent{}Graphically the generating function of a multi-$j$-symbol 
is represented by the graph of the multi-$j$-symbol with the internal 
lines labeled $A$, $B$, $C$, {\dots} and the external lines labeled 
($A$, $ \bar{A}$), ($B$, $ \bar{B}$), ($C$, $ \bar{C}$), {\dots} 
.

To derive the generating functions of the multi-$j$-symbol we have 
to introduce the following definition.

\pagebreak[3]

\noindent {\bf Definition} 2.1.

A curve $\omega $ running through a graph $\Gamma $ (following the 
lines of the graph) leads to the product $ P(\omega )$ of the variables 
of all lines that the curve passes. This curve may start and end 
in external lines. In this case the curve has an orientation indicated 
by an arrow. The external line where the curve starts from is represented 
by the unbared variable, whereas the terminal (external) line enters 
the product by its bared variable. If the curve is closed it has 
no external lines and no bared variables occur. Curves that start 
from or end in internal lines are not regarded.

Moreover $ P(\omega )$ is endowed with a sign: It picks up a minus 
sign 
\begin{enumerate}\item{}for every time it passes a line against the 
orientation of its arrow, 
\item{}for every time the direct way through a vertex (without crossing 
the third leg of the vertex) is a clockwise rotation, and 
\item{}it has an over all minus sign.
\end{enumerate} 
For sets $\Omega $ of curves $\omega $ we define 
\begin{equation}
\label{16}P\left( \Omega \right) =\prod _{\omega \in \Omega }P
\left( \omega \right) \hspace{.6ex},\hspace{2ex}\hbox
{\hspace{.38ex}if }\Omega \neq \emptyset \hbox{\hspace{.38ex}\hspace*{1cm}and
\hspace*{1cm}}P\left( \emptyset \right) =1\hspace{.6ex}.
\end{equation}

\vspace{1ex}
\noindent{}\centerline{\epsfbox{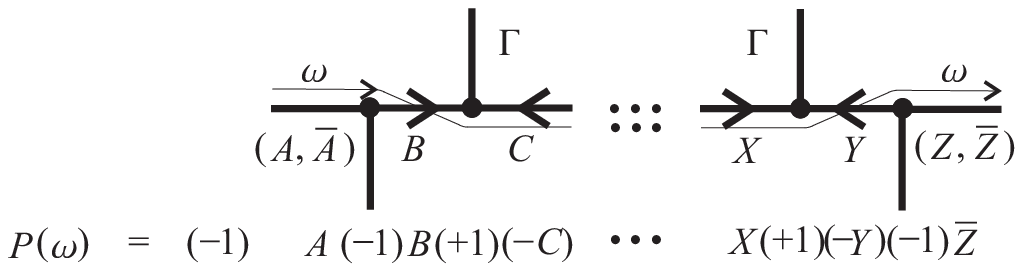}}

\vspace{1ex}
\noindent{}Note that for closed loops the direction we run through 
the loop is irrelevant: If we change the direction we pick up a 
phase $ (-1)^{{\#}_{\rm lines\hspace{.38ex}}}(-1)^{{\#}_{\rm vertices\hspace{.38ex}}
}=1$. The first factor comes form running through the lines in opposite 
direction and the second factor stems from the fact that we reverse 
the orientation we run through the vertices when we reverse the 
direction of the loop. Rule 3.\ in the above definition means for 
sets $\Omega $ of curves that $ P(\Omega )$ gets an over all sign 
$ (-1)^{{\rm {\#} \ connected \ components\hspace{.38ex}}}$.

\vspace{1ex}
\noindent{}Now, we can give graphical notations for the generating 
function of the last section: 
\begin{eqnarray}
&&\sum _{\textstyle  {abc\atop \alpha \beta \gamma }}\left( 
\begin{array}{ccc}a&b&c\\ 
\alpha &\beta &\gamma 
\end{array}\right) \frac{A^{a+\alpha }\bar{A}^{a-\alpha }B^{b+
\beta }\bar{B}^{b-\beta }C^{c+\gamma }\bar{C}^{c-\gamma }}{
\Delta \left( a,b,c\right) \sqrt{\left( a+\alpha \right) !
\left( a-\alpha \right) !\left( b+\beta \right) !\left( b-
\beta \right) !\left( c+\gamma \right) !\left( c-\gamma 
\right) !}}\nonumber \\ 
\label{4}&&=\exp\left( -\raisebox{-1.1cm}{$ \epsfbox{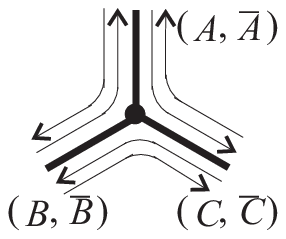
}$}\right)  \hspace{.6ex}.\\ 
&&\sum _{\textstyle  {abc\atop \alpha \beta \gamma }}\left( 
\begin{array}{ccc}a&b&c\\ 
\alpha &\beta &\gamma 
\end{array}\right) \sqrt\frac{\left( a-\alpha \right) !\left( b-
\beta \right) !\left( c-\gamma \right) !}{\left( a+\alpha 
\right) !\left( b+\beta \right) !\left( c+\gamma \right) !} 
\frac{A^{a+\alpha }\bar{A}^{a-\alpha }B^{b+\beta }\bar{B}^{b-
\beta }C^{c+\gamma }\bar{C}^{c-\gamma }}{\Delta \left( a,b,c
\right) }\nonumber \\ 
\label{12}&&=\frac{1}{1+\raisebox{-0.4cm}{$ \epsfbox{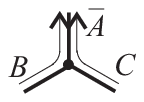
}$}} \hspace*{1ex}\frac{1}{1+\raisebox{-0.4cm}{$ \epsfbox{genfu04
.eps}$}} \hspace*{1ex}\frac{1}{1+\raisebox{-0.4cm}{$ \epsfbox
{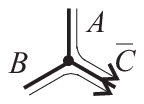}$}} \hspace{.6ex}.\\ 
&&\sum _{\textstyle  {abc\atop def}}\left\{ 
\begin{array}{ccc}a&b&c\\ 
d&e&f
\end{array}\right\} \frac{A^{2a}B^{2b}C^{2c}D^{2d}E^{2e}F^{2f}}{
\prod \limits _{\textstyle  {{\rm vertices \ }\atop v_{1},v_{2}
,v_{3}\in a,b,{\ldots},f}}\Delta \left( v_{1},v_{2},v_{3}
\right) }=\left( 1+\sum _{\textstyle  {{\rm non-overlapping\hspace{.38ex}}
\atop {\rm closed \ loops \ }\omega }}P\left( \omega \right) 
\right) ^{-2}{}\\ 
&&\sum _{\textstyle  {\textstyle  {abc\atop def}\atop ghk}}
\left\{ 
\begin{array}{ccc}a&b&c\\ 
d&e&f\\ 
g&h&k
\end{array}\right\} \frac{A^{2a}B^{2b}C^{2c}D^{2d}E^{2e}F^{2f}G^
{2g}H^{2h}K^{2k}}{\prod \limits _{\textstyle  {{\rm vertices \ }
\atop v_{1},v_{2},v_{3}\in a,b,{\ldots},f}}\Delta \left( v_{1},v
_{2},v_{3}\right) }=\left( 1+\!\!\sum _{\textstyle  {{\rm non-overlapping\hspace{.38ex}}
\atop {\rm closed \ loops \ }\omega }}\!\!P\left( \omega 
\right) \right) ^{-2}\!.\hspace*{2ex}
\end{eqnarray} 
The generating function is normalized by the square roots $ 
\Delta (v_{1},v_{2},v_{3})$ (Eq.\ (\ref{9})) for all vertices $ v
_{1},v_{2},v_{3}$ in $\Gamma $. The sum over all non-overlapping 
closed loops means the sum over all closed loops $\omega $ that 
pass each line at most once. (In the next section we will see that 
in general we have to deal with sets of loops which may have more 
than one connected component.) The polynomial $ P(\omega )$ is of 
order one in each variable.

\vspace{1ex}
\noindent{}\section{The Theorem\label{s3}} 
We start from Eq.\ (\ref{12}) which represents the building block 
for a general multi-$j$-symbol.

\vspace{1ex}
\noindent{}A generating function $ F(A_{1},\bar{A}_{1},A_{2},
\bar{A}_{2})$ with two external lines $ (A_{1},\bar{A}_{1})$ and 
$ (A_{2},\bar{A}_{2})$ is glued together to yield a generating function 
$ \tilde{F}(A_{12})$ with an internal line $A$$_{12}$ (running from 
$A$$_{1}$ to $A$$_{2}$) by the following procedure: 
\begin{equation}
\label{11}\tilde{F}\left( A_{12}\right) =\oint _{\partial U}
\frac{dc_{1}}{2\pi ic_{1}{}} \oint _{\partial U}\frac{dc_{2}}{2
\pi ic_{2}{}} F\left( -A_{12}c_{1},c_{2}^{-1},A_{12}c_{2},c_{1}^{
-1}\right) \hspace{.6ex}.
\end{equation} 
The integrals on the right hand side are loop integrals around the 
unit circle $ \partial U$ of the complex plane. The integrals may 
be evaluated using the residue theorem assuming $A$$_{12}$ (being 
an expansion parameter) is small.

It is readily checked that Eqs.\ (\ref{10}) and (\ref{11}) are equivalent: 
\begin{eqnarray*}
&&\oint _{\partial U}\frac{dc_{1}}{2\pi ic_{1}{}} \oint _{
\partial U}\frac{dc_{2}}{2\pi ic_{2}{}} F\left( -A_{12}c_{1},c_{
2}^{-1},A_{12}c_{2},c_{1}^{-1}\right) \\ 
&=&\sum _{a_{1}\alpha _{1}a_{2}\alpha _{2}}F_{a_{1}\alpha _{1}a_{
2}\alpha _{2}}\oint _{\partial U}\frac{dc_{1}}{2\pi ic_{1}{}} 
\oint _{\partial U}\frac{dc_{2}}{2\pi ic_{2}{}} \left( -A_{12}c_{
1}\right) ^{a_{1}+\alpha _{1}}{} \left( c_{2}\right) ^{-a_{1}+
\alpha _{1}}{} \left( A_{12}c_{2}\right) ^{a_{2}+\alpha _{2}}{} 
\left( c_{1}\right) ^{-a_{2}+\alpha _{2}}\\ 
&=&\sum _{a_{1}\alpha _{1}a_{2}\alpha _{2}}F_{a_{1}\alpha _{1}a_{
2}\alpha _{2}}\left( -1\right) ^{a_{1}+\alpha _{1}}A_{12}^{a_{1
}+\alpha _{1}+a_{2}+\alpha _{2}}\delta _{a_{1}+\alpha _{1},a_{2
}-\alpha _{2}}\delta _{a_{1}-\alpha _{1},a_{2}+\alpha _{2}}\\ 
&=&\sum _{a_{1}\alpha _{1}a_{2}\alpha _{2}}F_{a_{1}\alpha _{1}a_{
2}\alpha _{2}}A_{12}^{2a_{12}}\left( -1\right) ^{a_{1}+\alpha _{
1}}\delta _{\alpha _{1},-\alpha _{2}}\delta _{a_{12},a_{1}}
\delta _{a_{12},a_{2}}\hspace{.6ex}.
\end{eqnarray*} 
Note that the identifications $ a_{1}=a_{2}$, $ \alpha _{1}=-
\alpha _{2}$ cancel the square roots $ \sqrt\frac{(a_{1}+
\alpha _{1})!}{(a_{1}-\alpha _{1})!} \sqrt\frac{(a_{2}+\alpha _{
2})!}{(a_{2}-\alpha _{2})!}$. 

As an example let us glue two 3-$j$-symbols to obtain a 5-$j$-symbol 
with four external lines: 
\begin{eqnarray*}
&&\sum _{\textstyle  {abcde\atop \alpha _{1}\beta \gamma \alpha _{
2}\delta \varepsilon }}\left( 
\begin{array}{ccc}a&b&c\\ 
\alpha _{1}&\beta &\gamma 
\end{array}\right) \left( 
\begin{array}{ccc}a&d&e\\ 
\alpha _{2}&\delta &\varepsilon 
\end{array}\right) {a\choose \alpha _{1},\alpha _{2}}\sqrt\frac{
\left( b-\beta \right) !\left( c-\gamma \right) !}{\left( b+
\beta \right) !\left( c+\gamma \right) !} \sqrt\frac{\left( d-
\delta \right) !\left( e-\varepsilon \right) !}{\left( d+
\delta \right) !\left( e+\varepsilon \right) !}\\ 
&&\cdot \frac{A^{2a}B^{b+\beta }\bar{B}^{b-\beta }C^{c+\gamma }
\bar{C}^{c-\gamma }D^{d+\delta }\bar{D}^{d-\delta }E^{e+
\varepsilon }\bar{E}^{e-\varepsilon }}{\Delta \left( a,b,c
\right) \Delta \left( a,d,e\right) }\\ 
&=&\oint _{\partial U}\frac{dc_{1}}{2\pi ic_{1}{}} \oint _{
\partial U}\frac{dc_{2}}{2\pi ic_{2}{}} \frac{1}{1+\left( B-C
\right) c_{2}^{-1}{}} \hspace*{1ex}\frac{1}{1+\left( C+Ac_{1}
\right) \bar{B}}\\ 
&&\cdot \frac{1}{1+\left( -Ac_{1}-B\right) \bar{C}} 
\hspace*{1ex}\frac{1}{1+\left( D-E\right) c_{1}^{-1}{}} 
\hspace*{1ex}\frac{1}{1+\left( E-Ac_{2}\right) \bar{D}} 
\hspace*{1ex}\frac{1}{1+\left( Ac_{2}-D\right) \bar{E}}\\ 
&=&\frac{1}{1+\left( C-A\left( D-E\right) \right) \bar{B}} 
\hspace*{1ex}\frac{1}{1+\left( A\left( D-E\right) -B\right) 
\bar{C}}\\ 
&&\frac{1}{1+\left( E+A\left( B-C\right) \right) \bar{D}} 
\hspace*{1ex}\frac{1}{1+\left( -A\left( B-C\right) -D\right) 
\bar{E}}\\ 
&=&\frac{1}{1+\raisebox{-0.4cm}{$ \epsfbox{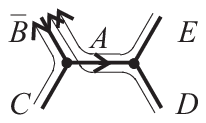}$}} 
\hspace*{1ex}\frac{1}{1+\raisebox{-0.4cm}{$ \epsfbox{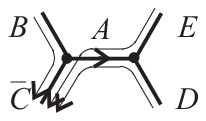
}$}} \hspace*{1ex}\frac{1}{1+\raisebox{-0.4cm}{$ \epsfbox{genfu09
.eps}$}} \hspace*{1ex}\frac{1}{1+\raisebox{-0.4cm}{$ \epsfbox
{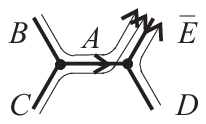}$}}\hspace{.6ex},
\end{eqnarray*} 
where we used the residue theorem and evaluated the residues inside 
the unit circle.

\vspace{1ex}
\noindent{}Before we formulate the general case we need some more 
notation.

\vspace{1ex}
\noindent{}{\bf Notation} 3.1.

\vspace{1ex}
\noindent{}Let $\Gamma $ be the graph of the multi-$j$-symbol $ S
(\Gamma )$ with external lines $ a_{j}$, $ j=1,2,{\ldots},J$ and 
internal lines $ a_{i}$, $ i=J+1,J+2,{\ldots},J+I$. If $ J=0$ then 
$\Gamma $ is closed.

Let $ L=I+J$ be the number of lines and $V$ the number of vertices 
in $\Gamma $.

We consider curves $\omega $ in $\Gamma $ as defined in Def.\ 2.1. 
A set $\Omega $ of curves in $\Gamma $ is said to run from $A$$
_{i}$ to $A$$_{j}$, $ i,j=1,{\ldots},J$ if one curve in $
\Omega $ is open and runs from $A$$_{i}$ to $A$$_{j}$ and all other 
curves in $\Omega $ are closed loops.

The degree deg$ _{A_{i}}(\omega )$ of a line $ A_{i}$ in a curve 
$\omega $ is the number of times $\omega $ passes through this line. 
The degree deg$ _{A_{i}}(\Omega )$ of a line $ A_{i}$ in a set $
\Omega $ of curves is the sum $ \sum _{\omega \in \Omega }\hbox
{\hspace{.38ex}deg\hspace{.38ex}}_{A_{i}}(\omega )$. A (set of) 
curve(s) $\omega $ ($\Omega $) is non-overlapping if deg$ _{A_{
i}}(\omega )\le 1\hspace*{1ex}\forall A_{i}\in \Gamma $ (deg$ _
{A_{i}}(\Omega )\le 1\hspace*{1ex}\forall A_{i}\in \Gamma $).

Let $\Omega _{ij}$, $ i,j=1,{\ldots},J$ be the set of all non-overlapping 
sets of curves in $\Gamma $ running from $A$$_{i}$ to $A$$_{j}$.

Let $\Omega _{0}$ be the set of all not overlapping sets of closed 
loops in $\Gamma $ (including the empty set $ \emptyset \in 
\Omega _{0}$).

Let $ P(\omega )$ ($ P(\Omega )$), as defined in Def.\ 2.1, be the 
sign endowed product of all lines $\omega $ ($\Omega $) runs through. 
We have deg$ _{A_{i}}(P(\Omega ))=\hbox{\hspace{.38ex}deg\hspace{.38ex}}_
{A_{i}}(\Omega )$.

\vspace{2ex}
\noindent{}We can draw some elementary conclusions from the fact 
that $\Gamma $ has only three-valent vertices.

\pagebreak[3]

\noindent {\bf Remark} 3.2.

\vspace{1ex}
\noindent{}1.\ $ J+2I=3V$. Moreover $ J=V+2\Longleftrightarrow 
\Gamma $ is a tree diagram.

\vspace{1ex}
\noindent{}2.\  
\begin{equation}
\label{13}|\Omega _{0}|=2^{I-V+1}=2^{\left( V+2-J\right) /2}
\hspace{.6ex}.
\end{equation} 
This is easily seen by induction. For $ J=V+2$ we obviously have 
$ \Omega _{0}=\{\emptyset \}$, $ |\Omega _{0}|=1$. If we reduce 
$J$ by two via gluing of two external lines then $ \Omega _{0,{\rm  
\ glued\hspace{.38ex}}}$ splits into two parts. One where the glued 
line has degree zero and one where it has degree one. The number 
of sets in the first part is obviously $ \Omega _{0,{\rm  \ unglued\hspace{.38ex}}
}$. The second part contains at least one loop. However there is 
a one to one correspondence between $ \Omega _{0,{\rm  \ unglued\hspace{.38ex}}
}$ and sets of curves running through the glued line: If we have 
two sets $\Omega _{1}$, $\Omega _{2}$ of non-overlapping curves 
with degree one at the glued vertex we can take the 'difference' 
of these sets by reducing the degrees of the lines in $ \Omega _{
1}\cup \Omega _{2}$ mod 2.\ This defines a set of loops in $ 
\Omega _{0,{\rm  \ unglued\hspace{.38ex}}}$. Thus $ |\Omega _{0
,{\rm  \ glued\hspace{.38ex}}}|=2\cdot |\Omega _{0,{\rm  \ unglued\hspace{.38ex}}
}|$.  

\vspace{1ex}
\noindent{}3.\  
\begin{equation}
\label{8}|\Omega _{ij}|=|\Omega _{0}|\hbox{ , if }J\ge 2
\hspace{.6ex}.
\end{equation} 
In deed, if we glue $A$$_{i}$ and $A$$_{j}$ we see that $ |
\Omega _{ij}|$ equals the number of sets in the glued graph with 
degree one at the glued line. This is $|\Omega _{0}|$, as explained 
above.\vspace{1cm}

\noindent{}
\pagebreak[3]

\noindent {\bf Theorem} 3.3. 
With the above notation and $ \Delta (a,b,c)$ as in Eq.\ (\ref{9}) 
we obtain a generating function for $ S(\Gamma )$ by 
\begin{eqnarray}
&&\sum _{^{a_{j}\alpha _{j},j\le J}_{a_{i},i>J}}S\left( \Gamma ,a
_{j},\alpha _{j},a_{i}\right) \frac{\prod \limits _{j\le J}A_{j
}^{a_{j}+\alpha _{j}}\bar{A}_{j}^{a_{j}-\alpha _{j}}\prod 
\limits _{i>J}A_{i}^{2a_{i}}}{\prod \limits _{\textstyle  {{\rm vertices 
\ }a_{k},a_{\ell },a_{m}\atop k,l,m=1,{\ldots},L}}\Delta 
\left( a_{k},a_{\ell },a_{m}\right) } \prod _{j\le J}\sqrt
\frac{\left( a_{j}-\alpha _{j}\right) !}{\left( a_{j}+\alpha _{
j}\right) !}\nonumber \\ 
\label{5}&=&\left( \sum _{\omega \in \Omega _{0}}P\left( \omega 
\right) \right) ^{|J|-2}\prod _{j\le J}\left( \sum _{
\textstyle  {\omega \in \Omega _{0}\cup \Omega _{ij}\atop i
\neq j}}P\left( \omega \right) \right) ^{-1}\hspace{.6ex}.
\end{eqnarray}

\pagebreak[3]

\noindent {\bf Proof}. 
We will prove the theorem in two steps.

First, we show the theorem is valid for tree graphs with $ 
\Omega _{0}=\{\emptyset \}$. This can easily be done by induction 
over the number of vertices in $\Gamma $. Obviously Eq.\ (\ref{5}) 
is valid for the 3-$j$-symbol, Eq.\ (\ref{14}). The gluing of further 
3-$j$-symbols follows closely the example of the 5-$j$-symbol in 
the previous section. Evaluating each loop integral amounts to a 
substitution and the orientation of the glued line is taken care 
of by the minus sign in the gluing prescription.

\vspace{1ex}
\noindent{}Second, we have to show that Eq.\ (\ref{5}) remains valid 
under gluing of any two external lines in $\Gamma $. The general 
result follows by induction over the number of times gluing is necessary. 
So, assume Eq.\ (\ref{5}) is valid and we want to glue (without 
restriction) $A$$_{1}$ and $A$$_{2}$. Let $ P_{0}(\Omega )$ be the 
$ A_{1},\bar{A}_{1},A_{2},\bar{A}_{2}$-independent part of $ P(
\Omega )$. The first factor on the right hand side of Eq.\ (\ref{5}) 
is independent of $ A_{1},\bar{A}_{1},A_{2},\bar{A}_{2}$. The second 
factor is 
\begin{eqnarray*}
&&\left( \sum _{\Omega \in \Omega _{0}}P\left( \Omega \right) +
\sum _{\Omega \in \Omega _{21}}P_{0}\left( \Omega \right) A_{2}
\bar{A}_{1}+\sum _{\Omega \in \Omega _{i1},i\neq 1,2}P_{0}
\left( \Omega \right) \bar{A}_{1}\right) ^{-1}{}\\ 
&\cdot &\left( \sum _{\Omega \in \Omega _{0}}P\left( \Omega 
\right) +\sum _{\Omega \in \Omega _{12}}P_{0}\left( \Omega 
\right) A_{1}\bar{A}_{2}+\sum _{\Omega \in \Omega _{i2},i\neq 1
,2}P_{0}\left( \Omega \right) \bar{A}_{2}\right) ^{-1}{}\\ 
&\cdot &\prod _{2<j\le J}\left( \sum _{\Omega \in \Omega _{0}}P
\left( \Omega \right) +\sum _{\Omega \in \Omega _{1j}}P_{0}
\left( \Omega \right) A_{1}+\sum _{\Omega \in \Omega _{2j}}P_{0
}\left( \Omega \right) A_{2}+\sum _{\textstyle  {\Omega \in 
\Omega _{ij}\atop i\neq 1,2,j}}P\left( \Omega \right) \right) ^{
-1}\hspace{.6ex}.
\end{eqnarray*} 
According to the gluing prescription, Eq.\ (\ref{11}), we now substitute 
$ A_{1}\rightarrow -A_{12}c_{1}$, $ \bar{A}_{1}\rightarrow c_{2
}^{-1}$, $ A_{2}\rightarrow A_{12}c_{2}$, $ \bar{A}_{2}
\rightarrow c_{1}^{-1}$ and multiply by $ \frac{1}{2\pi ic_{1}
{}} \frac{1}{2\pi ic_{2}}$. The loop integrals over $c$$_{1}$ and 
$c$$_{2}$ amount to substituting the first and the second factor 
into the product over $ j>2$ (we assume all $A$$_{i}$, $A$$_{j}$ 
are small and evaluate the residues inside the unit circle). To 
be precise we obtain 
\begin{eqnarray*}
&&\hspace{-0.7cm}\left( \sum _{\Omega \in \Omega _{0}}P\left( 
\Omega \right) +\sum _{\Omega \in \Omega _{21}}A_{12}P_{0}
\left( \Omega \right) \right) ^{-1}\left( \sum _{\Omega \in 
\Omega _{0}}P\left( \Omega \right) -\sum _{\Omega \in \Omega _{
12}}A_{12}P_{0}\left( \Omega \right) \right) ^{-1}\cdot \prod _
{2<j\le J}\\ 
&&\hspace{-0.7cm}\left( \sum _{\Omega \in \Omega _{0}}\!P\left( 
\Omega \right) +\!\!\!\sum _{\textstyle  {\Omega \in \Omega _{i
j}\atop i\neq 1,2,j}}\!P\left( \Omega \right) +\frac{\sum 
\limits _{\Omega \in \Omega _{1j}}A_{12}P_{0}\left( \Omega 
\right) \sum \limits _{\textstyle  {\Omega \in \Omega _{i2}
\atop i\neq 1,2}}P_{0}\left( \Omega \right) }{\sum \limits _{
\Omega \in \Omega _{0}}P\left( \Omega \right) -\sum \limits _{
\Omega \in \Omega _{12}}A_{12}P_{0}\left( \Omega \right) } -
\frac{\sum \limits _{\Omega \in \Omega _{2j}}A_{12}P_{0}\left( 
\Omega \right) \sum \limits _{\textstyle  {\Omega \in \Omega _{
i1}\atop i\neq 1,2}}P_{0}\left( \Omega \right) }{\sum \limits _
{\Omega \in \Omega _{0}}P\left( \Omega \right) +\sum \limits _{
\Omega \in \Omega _{21}}A_{12}P_{0}\left( \Omega \right) } 
\right) ^{-1}\!.
\end{eqnarray*} 
If we reverse the direction of the open curve in $\Omega $ we obtain 
$ \sum _{\Omega \in \Omega _{21}}P_{0}(\Omega )=-\sum _{\Omega 
\in \Omega _{12}}P_{0}(\Omega )$. We denote the glued graph with 
$ \Gamma '$ and the set of all non-overlapping closed loops in $ 
\Gamma '$ with $ \Omega _{0}'$. Analogously we define $ \Omega _{
ij}'$ as the set of non-overlapping sets of curves in $ \Gamma '$ 
running from $A$$_{i}$ to $A$$_{j}$. Thus 
\begin{displaymath}
\sum _{\Omega \in \Omega _{0}}P\left( \Omega \right) +\sum _{
\Omega \in \Omega _{21}}A_{12}P_{0}\left( \Omega \right) =\sum _
{A_{12}\notin \Omega \in \Omega _{0}'}P\left( \Omega \right) +
\sum _{A_{12}\in \Omega \in \Omega _{0}'}P\left( \Omega 
\right) =\sum _{\Omega \in \Omega _{0}'}P\left( \Omega \right) 
\hspace{.6ex}.
\end{displaymath} 
This simplifies the above result to 
\begin{eqnarray*}
&&\hspace{-0.7cm}\left( \sum _{\Omega \in \Omega _{0}'}P\left( 
\Omega \right) \right) ^{J-4}\prod _{2<j\le J}\left( \left( 
\sum _{\Omega \in \Omega _{0}}P\left( \Omega \right) \right) 
\left( \sum _{\Omega \in \Omega _{0}'}P\left( \Omega \right) +
\sum _{\textstyle  {\Omega \in \Omega _{ij}\atop i\neq 1,2,j}}P
\left( \Omega \right) \right) \right. \\ 
&&\hspace{-0.7cm}\left. +\!\sum _{A_{12}\in \Omega \in \Omega _{
0}'}\!P\left( \Omega \right) \!\sum _{\textstyle  {\Omega \in 
\Omega _{ij}\atop i\neq 1,2,j}}\!P\left( \Omega \right) +\!
\sum _{\Omega \in \Omega _{1j}}\!A_{12}P_{0}\left( \Omega 
\right) \!\sum _{\textstyle  {\Omega \in \Omega _{i2}\atop i
\neq 1,2}}\!P_{0}\left( \Omega \right) -\!\sum _{\Omega \in 
\Omega _{2j}}\!A_{12}P_{0}\left( \Omega \right) \!\sum _{
\textstyle  {\Omega \in \Omega _{i1}\atop i\neq 1,2}}\!P_{0}
\left( \Omega \right) \right) ^{-1}.
\end{eqnarray*} 
We need the following notation:

Let $ \Omega _{i12}'$ be the set of non-overlapping curves in $ 
\Gamma '$ running from $A$$_{i}$ to $A$$_{12}$ parallel to the orientation 
of $A$$_{12}$.

Let $ \Omega _{i21}'$ be the set of non-overlapping curves in $ 
\Gamma '$ running from $A$$_{i}$ to $A$$_{12}$ anti-parallel to 
the orientation of $A$$_{12}$.

Let $ \Omega _{12j}'$ be the set of non-overlapping curves in $ 
\Gamma '$ running from the $A$$_{2}$ vertex of $A$$_{12}$ to $A$
$_{j}$ without passing through $A$$_{12}$.

Let $ \Omega _{21j}'$ be the set of non-overlapping curves in $ 
\Gamma '$ running from the $A$$_{1}$ vertex of $A$$_{12}$ to $A$
$_{j}$ without passing through $A$$_{12}$.

\vspace{1ex}
\noindent{}With these notations $ \Omega _{i12}'\cup \Omega _{12
j}'$ contains a curve running from $A$$_{i}$ to $A$$_{j}$ passing 
$A$$_{12}$ parallel to the orientation of $A$$_{12}$, and $ 
\Omega _{i21}'\cup \Omega _{21j}'$ contains a curve running from 
$A$$_{i}$ to $A$$_{j}$ passing $A$$_{12}$ anti-parallel to the orientation 
of $A$$_{12}$. Now the last three terms can be written as 
\begin{displaymath}
\sum _{\textstyle  {A_{12}\in \Omega _{1}\in \Omega _{0}'\atop A_{
12}\notin \Omega _{2}\in \Omega _{ij}',i\neq 1,2,j}}P\left( 
\Omega _{1}\cup \Omega _{2}\right) +\sum _{\textstyle  {\Omega _{
1}\in \Omega _{i21}',i\neq 1,2\atop \Omega _{2}\in \Omega _{21j
}'}}P_{0}\left( \Omega _{1}\cup \Omega _{2}\right) +\sum _{
\textstyle  {\Omega _{1}\in \Omega _{i12}',i\neq 1,2\atop 
\Omega _{2}\in \Omega _{12j}'}}P_{0}\left( \Omega _{1}\cup 
\Omega _{2}\right) =X\hspace{.6ex}.
\end{displaymath} 
To get the signs right in this expression we have to bear in mind 
that in the last two terms we gain a minus sign by connecting the 
open curve in $\Omega _{1}$ and $\Omega _{2}$ and in the middle 
term we get another minus sign by running against the orientation 
of $A$$_{12}$. Now we can simplify $X$ as follows: Whenever the 
degree of $ \Omega _{1}\cup \Omega _{2}$ is 2 at some line $A$$
_{k}$ we can swap the end points of the lines to get a crossing 
instead of to parallel lines (and vice versa). This gives a new 
set $ \overline {\Omega _{1}\cup \Omega _{2}}$. We find $ P(
\Omega _{1}\cup \Omega _{2})=-P(\overline {\Omega _{1}\cup 
\Omega _{2}})$, where the minus sign stems from either the change 
of orientations if we started from two anti-parallel lines or from 
gaining or losing a connected component if we started from parallel 
lines. A non-trivial but purely geometrical calculation leads to 
the expression 
\begin{displaymath}
X=\sum _{\Omega \in \Omega _{0}}P\left( \Omega \right) \sum _{
\textstyle  {A_{12}\in \Omega \in \Omega _{ij}'\atop i\neq 1,2,j
}}P\left( \Omega \right) \hspace{.6ex}.
\end{displaymath} 
Altogether we obtain 
\begin{displaymath}
\left( \sum _{\Omega \in \Omega _{0}}P\left( \Omega \right) 
\right) ^{-J+2}\left( \sum _{\Omega \in \Omega _{0}'}P\left( 
\Omega \right) \right) ^{J-4}\!\!\prod _{2<j\le J}\left( \sum _
{\Omega \in \Omega _{0}'}P\left( \Omega \right) +\!\sum _{
\textstyle  {A_{12}\notin \Omega \in \Omega _{ij}'\atop i\neq 1
,2,j}}P\left( \Omega \right) +\!\sum _{\textstyle  {A_{12}\in 
\Omega \in \Omega _{ij}'\atop i\neq 1,2,j}}P\left( \Omega 
\right) \right) \hspace{.6ex}.
\end{displaymath} 
The first factor cancels the prefactor in Eq.\ (\ref{5}). The sums 
combine to $ \sum \limits _{\Omega \in \Omega _{0}'\cup \Omega _{
ij}',i\neq 1,2,j}P(\Omega )$, which establishes the desired result.\hfill $\Box $
\pagebreak[3] 

\vspace{1ex}
\noindent{}It is possible to rewrite Eq.\ (\ref{5}) in the spirit 
of Eq.\ (\ref{1}) which makes it slightly more symmetric.

\vspace{1ex}
\noindent{}
\pagebreak[3]

\noindent {\bf Corollary} 3.4. 
\begin{eqnarray}
&&\sum _{\textstyle  {a_{j}\alpha _{j},j\le J\atop a_{i},i>J}}S
\left( \Gamma ,a_{j},\alpha _{j},a_{i}\right) \frac{\prod 
\limits _{j\le J}A_{j}^{a_{j}+\alpha _{j}}\bar{A}_{j}^{a_{j}-
\alpha _{j}}\prod \limits _{i>J}A_{i}^{2a_{i}}}{\prod \limits _
{\textstyle  {{\rm vertices \ }a_{k},a_{\ell },a_{m}\atop k,l,m
=1,{\ldots},L}}\Delta \left( a_{k},a_{\ell },a_{m}\right) } 
\prod _{j\le J}\frac{1}{\sqrt{\left( a_{j}-\alpha _{j}\right) !
\left( a_{j}+\alpha _{j}\right) !}} \nonumber \\ 
\label{6}&=&\left( \sum _{\omega \in \Omega _{0}}P\left( \omega 
\right) \right) ^{-2}\exp\left( -\frac{\sum _{\omega \in 
\Omega _{ij},i\neq j}P\left( \omega \right) }{\sum _{\omega 
\in \Omega _{0}}P\left( \omega \right) }\right) \hspace{.6ex}.
\end{eqnarray}

\pagebreak[3]

\noindent {\bf Proof}. 
We proceed in the same way as we did when we derived Eq.\ (\ref{14}) 
from Eq.\ (\ref{1}). We substitute $ \bar{A}_{i}\rightarrow t_{
i}\bar{A}_{i}$ $ \forall i>J$ in Eq.\ (\ref{6}) and multiply both 
sides by exp$ (-\sum _{i>J}t_{i})$. Integrating over the $ t_{i
}$ from 0 to $\infty $ yields Eq.\ (\ref{5}). This proves the corollary 
since the transformation between both generating functions is obviously 
invertible.\hfill $\Box $
\pagebreak[3] 

\section{Results and Outlook} 
We found explicit geometric results for generating functions of multi-$j$-symbols. 
This result provides closed expressions for the multi-$j$-symbols 
themselves in terms of finite sums. The right hand side of Eq.\ 
(\ref{5}) has the form 
\begin{equation}
\left( 1+A\right) ^{J-2}\prod _{j=1}^J\left( 1+A+B_{j}\right) ^{
-1}\hspace{.6ex},\hspace{2ex}A=\sum _{\emptyset \neq \Omega 
\in \Omega _{0}}P\left( \Omega \right) 
\hspace{.6ex},\hspace{2ex}B=\sum _{\Omega \in \Omega _{ij},i
\neq j}P\left( \Omega \right) \hspace{.6ex}.
\end{equation} 
This may be expanded as 
\begin{equation}
\sum _{k_{1}=0}^{\infty }\cdots \sum _{k_J=0}^{\infty }\left( -B_{
j}\right) ^{k_{j}}\left( 1+A\right) ^{-\sum _{j}k_{j}-2}=\sum _
{k_{1}=0}^{\infty }\cdots \sum _{k_J=0}^{\infty }\left( -B_{j}
\right) ^{k_{j}}\sum _{k_{0}=0}^{\infty }{\sum _{j}k_{j}+k_{0}+1
\choose k_{0}}A^{k_{0}}\hspace{.6ex}.
\end{equation} 
Expanding $A$ and $B$$_{j}$ yields $ J(J-1)2^{(V+2-J)/2}+2^{(V+2
-J)/2}-1$ sums (Eqs.\ (\ref{13}), (\ref{8})). Comparing coefficients 
gives $ I+2J$ relations. This means that a full expansion provides 
an expression for the multi-$j$-symbols in terms of $ (J^{2}-J+1
)2^{(V+2-J)/2}-3(V+J)/2-\delta _{J,0}$ finite sums. The Kronecker 
delta reflects the fact that, if $ J\neq 0$, the sum over all magnetic 
quantum numbers is zero automatically; this identity does not reduce 
the number of independent sums. If we specify to the 3-$j$-symbol 
$ V=1$, $ J=3$, or to the 6-$j$-symbol $ V=4$, $ J=0$ we obtain 
single sums which are the well known results \cite{A} 
\begin{eqnarray}
\left( 
\begin{array}{ccc}a&b&c\\ 
\alpha &\beta &\gamma 
\end{array}\right) &=&\Delta \left( a,b,c\right) \sqrt{\left( a+
\alpha \right) !\left( a-\alpha \right) !\left( b+\beta 
\right) !\left( b-\beta \right) !\left( c+\gamma \right) !
\left( c-\gamma \right) !}\nonumber \\ 
&&\hspace{-2cm} \sum _{z}\frac{\left( -1\right) ^{z+a-b-\gamma }
\delta _{\alpha +\beta +\gamma ,0}}{z!\left( a+b-c-z\right) !
\left( a-\alpha -z\right) !\left( b+\beta -z\right) !\left( c-b
+\alpha +z\right) !\left( c-a-\beta +z\right) !}\\ 
\left\{ 
\begin{array}{ccc}a&b&c\\ 
d&e&f
\end{array}\right\} &=&\Delta \left( a,b,c\right) \Delta \left( a
,e,f\right) \Delta \left( b,d,f\right) \Delta \left( c,d,e
\right) \nonumber \\ 
&&\hspace{-2cm} \sum _{z}\frac{\left( -1\right) ^{z}\left( z+1
\right) !}{\left( z-a-b-c\right) !\left( z-a-e-f\right) !
\left( z-b-d-f\right) !\left( z-c-d-e\right) !}\nonumber \\ 
&&\hspace{-1.5cm} \cdot \hspace*{1ex}\frac{1}{\left( a+b+d+e-z
\right) !\left( b+c+e+f-z\right) !\left( a+c+d+f-z\right) !} 
\hspace*{2ex}\hspace{.6ex}.
\end{eqnarray} 
The last equation has been derived from the first equation by Racah 
\cite{C} with quite tedious calculations.

\vspace{1ex}
\noindent{}The analogous formula for the 9-$j$-symbol contains already 
a six-fold finite sum and is therefore hardly of practical use. 
However the expansion of the generating function allows us to give 
a geometrical interpretation of the multi-$j$-symbols themselves. 
Let us for simplicity stick to the closed case $ J=0$. The angular 
momenta $a$$_{1}$, {\dots}, $ a_L$ label the lines of the graph 
$\Gamma $. The multi-$j$-symbol counts the number of different ways 
how non-overlapping sets of loops can be laid on top of each other 
so that the total degree of each line $k$ is given by $a$$_{k}$. 
Each possible solution is weighted by one plus the number of layers 
needed. In addition it has a combinatorial factor of how many different 
ways the layers can be permuted.

The search for further applications of these results is not yet completed. 
It may help to evaluate multi loop Feynman diagrams since the result 
of the angular integrations is given by the square of a multi-$j$-symbol 
(more than three-valent vertices are blown up to a chain of three-valent 
vertices to give the graph of a multi-$j$-symbol).

\vspace{1ex}
\noindent{}Obvious generalizations are the application to higher 
rank Lie-groups or to quantum groups. These may be covered in future 
publications. 
\section*{Aknowledgement} 
I am very grateful to Prof.\ O'Raifeartaigh for his kind encouragement 
and valuable discussions. 
\end{document}